\documentclass[preprint,aip,cha,amsfonts,amssymb,amsmath,notitlepage]{revtex4-1} 

\usepackage{graphicx}
\usepackage{siunitx}
\usepackage{dsfont}
\usepackage{xcolor}
\usepackage{bm}
\usepackage{physics}
\usepackage[utf8]{inputenc}
\usepackage{subfigure}
\usepackage{hyperref}
\usepackage{mathtools}

\newcommand{\du}[2][]{\mathop{\mathrm{d}^{#1}#2}}
\newcommand{\vv}[1]{\bm{#1}}
\newcommand{\ee}{\mathrm{e}}
\newcommand{\ii}{\mathrm{i}}
\newcommand{\modu}[2][]{{\left|#2\right|}^{#1}}
\newcommand{\tpsi}{\tilde{\psi}}
\newcommand{\ma}{\modu[2]{A_0}}
\graphicspath{{./figures/}}

\begin{document}

\title{Convergent measure of focal extent, and largest peak intensity for non-paraxial beams}
\author{Petar Andrejić}
\affiliation{School of Chemical and Physical Sciences, Victoria University of Wellington, PO Box 600, Wellington, New Zealand}
\date{\today}

\begin{abstract}
  Second moment beam widths are commonly used in paraxial optics to define the focal extent of beams. However, second moments of arbitrary beams are not guaranteed to be finite. I propose the \emph{focal concentration area} as a measure of beam focal area, defined to be the ratio of total radial intensity to radial intensity regulated by a unit area Gaussian distribution. I use the Dirac delta limit of this distribution to establish a rigorous upper bound on the peak intensity of non-paraxial beams of a given total intensity, and show that this is achieved by the recently proposed `proto-beam' solution. I discuss the generalisation to electromagnetic beams, and find the same lower bound as the scalar case. This bound cannot be achieved for any physical beam, and as such the physical lower bound must be higher.
\end{abstract}

\maketitle
\section{Introduction}\label{sec:introduction}
It is of great experimental and theoretical importance to be able to define the focal extent of beams. By focal extent, I refer to the \emph{degree of localisation of intensity to a given point}. Beams arise as monochromatic solutions of wavelike equations in multiple contexts, such as electromagnetism, acoustics, and quantum mechanics. Monochromatic waves in all of these contexts have a spatial dependence given by solutions to the Helmholtz equation,
\begin{equation}
  \nabla^2\psi +k^2\psi=0,\quad k=k(\omega),
\end{equation}
with the full solution of the original wavelike equation being given by $\psi(\vv{r})\ee^{-i\omega t}$. The modulus squared of $\psi$ has dimensions of intensity per unit area. It is common to instead work with a dimensionless normalised wavefunction, $\tpsi$, so that $\psi=A_0\tpsi$, for some complex amplitude $A_0$ with the same dimension as $\psi$.

The direction of the beams propagation is referred to as the `beam axis', and is commonly taken to be $z$. It is natural to write the beam as an envelope times a plane wave propagating along the beam axis, $\psi(\vv{r})=u(\vv{r})\ee^{ikz}$. A common assumption, valid for many beams used in experiment, is that the envelope varies over a far larger length scale than the wavelength, $\lambda\equiv 2\pi /k$. This results in the approximate differential equation for the envelope,
\begin{equation}
  {\nabla_\perp}^2 u + 2ik u \approx 0,\quad \pdv[2]{u}{z}\ll k\pdv{u}{z}.
\end{equation}
This is referred to as the paraxial Helmholtz equation, and is solved by the commonly encountered Gaussian solutions, which have a fundamental mode (in complex point source notation, cylindrical coordinates) given by
\begin{equation}
  \tpsi_G(\rho,z)=\frac{b}{b+\ii z}\exp\left(ikz-\frac{k\rho^2}{2(b+\ii z)}\right).
\end{equation}
In general, the focal extent of the commonly encountered paraxial beams is simple to define. For centrally peaked beam profiles, such as the Gaussian beam, it is common to use shape parameters such as spot size or beam radius as a focal measure~\cite[sec.~3.1]{SalehFundamentalsphotonics2001}\cite[sec.~3.2.1]{NovotnyPrinciplesnanooptics2012}\cite[sec. 17.1]{SiegmanLasers1986}. For the Gaussian beam, the beam radius is defined as the radius at which the intensity, $\modu[2]{\psi}$, has decayed to $\ee^{-2}$ of the peak value, and is given by $w=\sqrt{2(b^2+z^2)/kb}$.

These parameters are more ambiguous for hollow beams, which can be resolved by using a more general measure, such as second moment integrals. A typical example of paraxial solutions that are hollow are the Laguerre-Gauss beams, which form a basis for paraxial beams. They are parametrised by two integers $m,p$, and defined as
\begin{equation}\label{eq:laguerre-gauss}
\begin{aligned}
	\tpsi_{0,m}(\rho,\phi,z)
	&=
	{\left(\frac{\rho}{b+\ii z}\right)}^{|m|}\ee^{\ii m \phi}\tpsi_G(\rho,z),
	\\
	\tpsi_{p,m}(\rho,\phi,z)&=\ee^{-2\ii p \arctan(z/b)}L_{p}^{|m|}\left(\frac{kb\rho^2}{b^2+z^2}\right)\tpsi_{0,m}(\rho,\phi,z).
\end{aligned}
\end{equation}
For $m\neq 0$, the beams are hollow, that is, have zero intensity along the beam axis. As such, defining the feature to use as the `spot' is far more ambiguous.

A generally applicable method of measuring focal extent have been standardised via ISO1146~\cite{iso11146}, which settles on the second moments of beam intensity as the standard measure of beam diameter. Specifically, this parameter defines the square of beam width to be twice the variance, the ratio of the second moment of intensity to the total intensity of a beam,
\begin{equation}
  {W}^2 = 2{\sigma_\rho}^2,\quad {\sigma_\rho}^2\equiv \frac{\int \rho^2\modu[2]{\psi}\du[2]{r}}{\int \modu[2]{\psi}\du[2]{r}}.
\end{equation}
For the Gaussian beam, this reproduces the previous spot size,
\begin{equation}
  W_G=\sqrt{\frac{2(b^2+z^2)}{kb}}=\rho_0
\end{equation}
For the Laguerre-Gauss beams, it can be shown that the beam width for index $p,m$ is given by~\cite{PhillipsSpotsizedivergence1983,Saghafibeampropagationfactor1998}
\begin{equation}
  {W_{p,m}}^2={W_G}^2(2p+|m|+1).
\end{equation}
Note how whatever the choice of parameters, the beam area $W^2$ increases quadratically with $z$. For paraxial beams, it is an \emph{exact} result that the beam area \emph{always} evolves quadratically with $z$~\cite{LaviGeneralizedbeamparameters1988,SiegmanNewdevelopmentslaser1990}, and this is one method of defining a `beam quality' parameter (also referred to as $M^2$)~\cite{SiegmanDefiningmeasuringoptimizing1993}. 
\begin{equation}\label{eq:quad}
  {W(z)}^2={W(z_0)}^2+M^2{(z-z_0)}^2,\quad M^2\geq 1.
\end{equation}
Equality is reached iff.~the beam is a Gaussian beam~\cite{SiegmanNewdevelopmentslaser1990}. It is clear from~\eqref{eq:quad} that the divergence angle of a particular beam is given by $\theta=\arctan(M/kW(z_0))$.

The issue with using the second moment as a beam width is that the second moment is not guaranteed to converge for arbitrary beams. For example, the Airy diffraction pattern, the result of Fraunhofer diffraction from a circular aperture, has an intensity pattern given by
\begin{equation}
  \tpsi_{Ai}(\vv{r})=\frac{J_1(ka\rho/r)}{ka\rho/r},\quad \rho=\sqrt{x^2+y^2}.
\end{equation}
The second moment of this intensity pattern is divergent, yet the total intensity is finite. In general, the conditions for a beam to be considered physical are the finiteness of its invariants; a resulting necessary condition is square integrability. The finiteness of the second moments however is not a necessity, and even in the paraxial regime it is well known that beams with sharp edged near field profiles result in diverging second moments in the far field. In Section~\ref{sec:focal-concentration}, I examine alternative methods of defining beam widths using regulators.

The above results have been all for paraxial beams. Paraxial beams as mathematical solutions can be focused to arbitrarily small focal extents. For example, one can simply make the parameter $b$ in the Gaussian beam arbitrarily close to 0. It is of course well known that real beams will eventually be diffraction limited; the Rayleigh limit gives an approximate lower bound of the `waist' of a beam, its lowest beam radius. One of the goals of this paper is to investigate the lower limits of focal extent for scalar beams. For this task, the paraxial approximation is insufficient, and we must use exact solutions of the Helmholtz equation.

\section{Limitations of spot size and second moment for exact solutions}\label{sec:limitations}
An arbitrary, forward propagating, exact solution of the Helmholtz equation can be expressed as a superposition of forward propagating Bessel beams,
\begin{widetext}
\begin{equation}\label{eq:expansion}
  \tpsi(\rho,\phi,z)=\frac{1}{k^2}\sum_{m\in\mathbb{Z}}\ee^{\ii m\phi}\int_0^k f_m(\kappa/k)J_m(\kappa\rho)\ee^{\ii q z}\kappa\du\kappa,\quad q^2+\kappa^2=k^2.
\end{equation}
\end{widetext}
Porras~\cite{Porrasbestqualityoptical1994} uses variational methods to find a the weight function $f_0(\kappa)$ that has the lowest second moment spot size. He shows that the beam
\begin{equation}\label{eq:porras-beam}
  \tpsi_P(\rho,z)=\frac{1}{k^2}\int_{0}^{k}J_0(\sqrt{b_0}\kappa/k)J_0(\kappa\rho)\ee^{\ii qz}\kappa\du\kappa,\quad \kappa^2+q^2=k^2,
\end{equation}
where $b_0$ is the first zero of $J_0$, has the smallest second moment possible for an exact solution of the Helmholtz equation.

However, as previously mentioned, it is not guaranteed that the beam will have a finite second moment, and as I shall argue in shortly, certain beams with divergent second moments can nevertheless have smaller focal extents.

For azimuthally symmetric beams, with approximately Gaussian profiles, one could attempt to use the decay of the beam profile to define a spot size, much as one can for a Gaussian. For spot size to be an adequate measure of focal extent, a significant portion, in the vicinity of 90\%, of the intensity should be contained within.

Within the $\ee^{-2}$ contour, we can calculate the fraction of total intensity of the Gaussian beam. The total intensity in a plane of constant $z$ is given by
\begin{equation}
  \frac{2\pi b^2}{b^2+z^2}\int_0^\infty \exp\left(-\frac{kb\rho^2}{b^2+z^2}\right)\rho\du{\rho}=\frac{\pi b}{k}\ma,
\end{equation}
whereas the intensity within the contour is given by
\begin{equation}
  \frac{2\pi b^2}{b^2+z^2}\ma\int_0^{\rho_0(z)} \exp\left(-\frac{kb\rho^2}{b^2+{z}^2}\right)\rho\du{\rho}=\frac{\pi b}{k}(1-\ee^{-2})\ma.
\end{equation}
Thus, approximately 86.5\% of the total intensity is contained in the initial spot, independent of the choice of $z$.

\begin{figure}[hbt]
\centering
\subfigure[\label{sfig:first-zero}]{\includegraphics[width=3in]{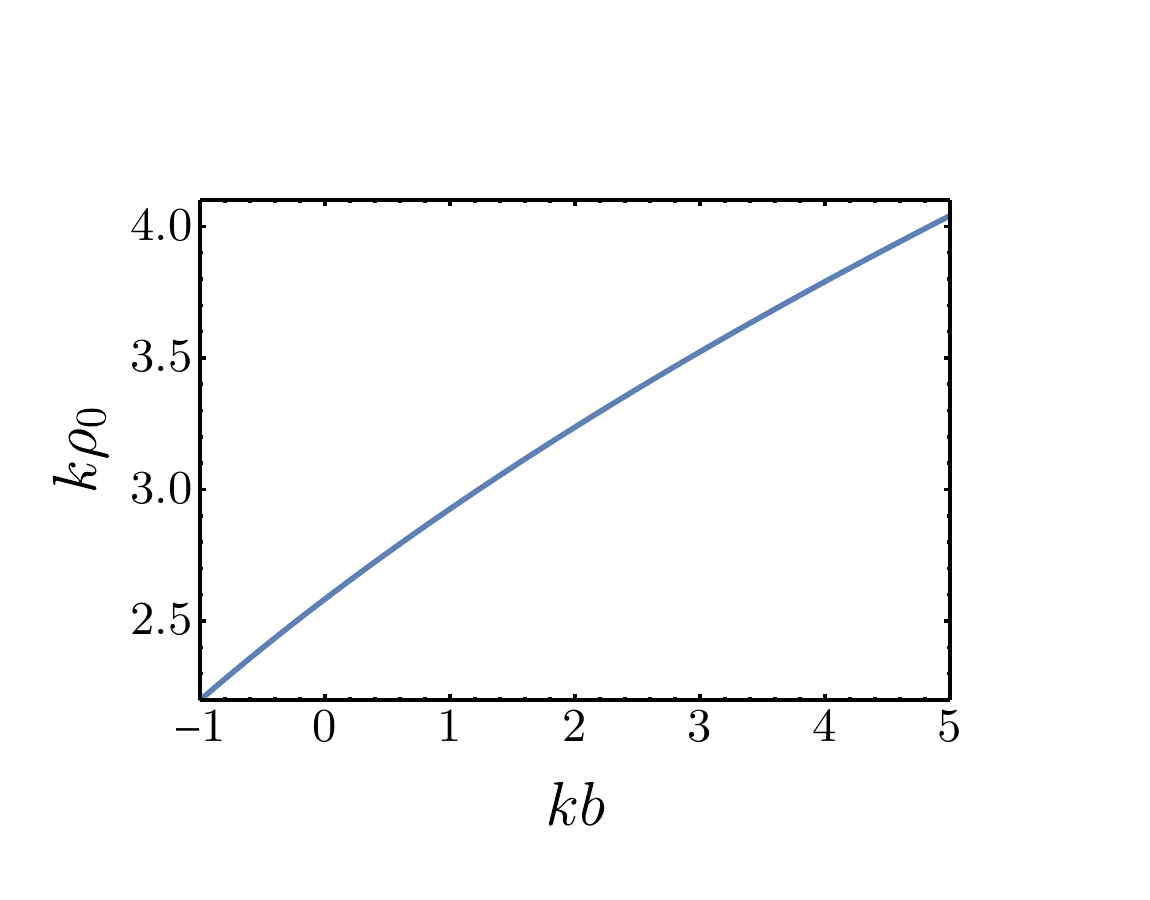}}
\subfigure[\label{sfig:fraction-total}]{\includegraphics[width=3in]{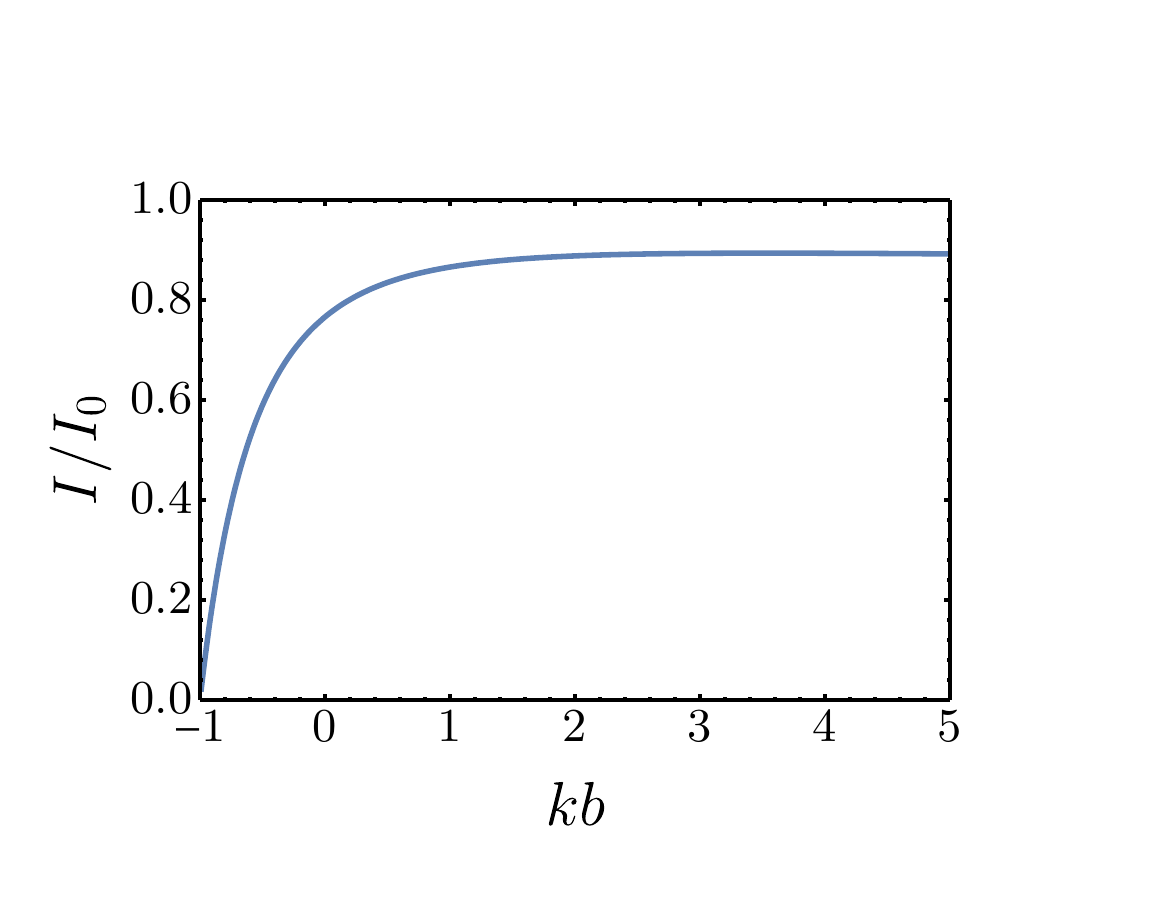}}
  \caption{ Radius of first zero,~\subref{sfig:first-zero}, and fraction of total intensity contained within this radius,~\subref{sfig:fraction-total}, for beam family given in~\eqref{eq:power-beam}. Note how below $kb\approx1$ the fraction of total intensity rapidly drops.}\label{fig:kb_intensity}
\end{figure}

\begin{figure}[hbt]
    \includegraphics[width=5in]{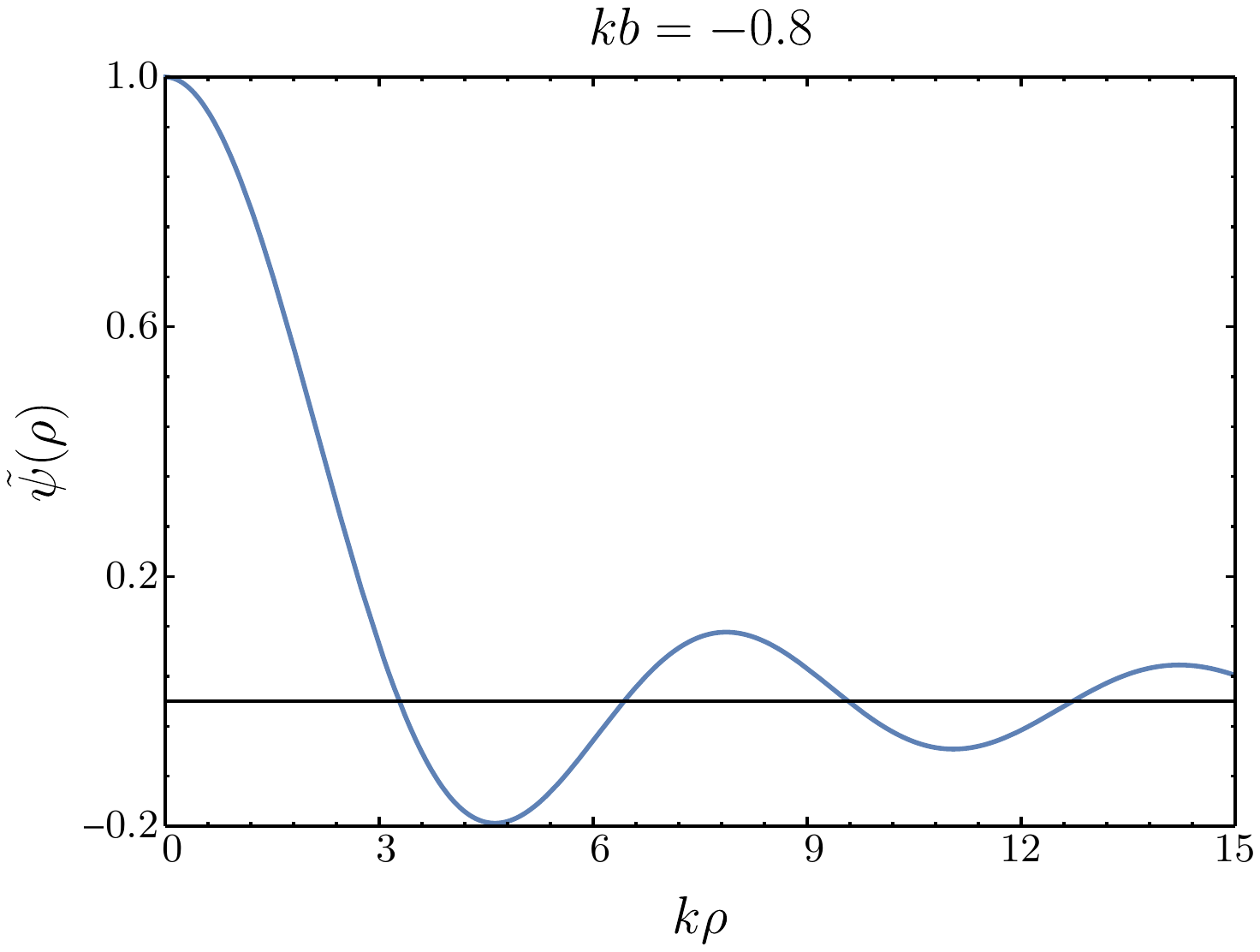}
  \caption{Focal plane wavefunction for the beam given by~\eqref{eq:power-beam}, with $kb=-0.8$. The beam is highly oscillatory, and decays slowly. As a result, the intensity contained in the initial spot, defined via the radius where intensity drops to $\ee^{-2}$ of peak, is only $33.4\%$ of total.}\label{fig:b0_2-beam}
\end{figure}

The Gaussian beam decays uniformly. Spot size becomes problematic for highly oscillatory decay, as is the case for diffraction limited beams. As an example, consider the family of beams with a power weight function
\begin{equation}\label{eq:power-beam}
  \tpsi(b,\rho,z)=\frac{kb+2}{k^{kb+2}}\int_0^k q^{kb} \ee^{\ii qz}J_0(\kappa\rho)q\du{q}, \quad kb>-1.
\end{equation}
The beam is normalised to have unit peak amplitude, and the condition $kb>1$ is necessary for the beam to be square integrable. In general, the wave-function does not have a closed form expression, however it is simple to evaluate numerically, via quadrature, or methods such as Levin integration~\cite{LevinFastintegrationrapidly1996a,LevinProcedurescomputingone1982}. For now let us restrict our attention to the focal plane $z=0$, where the beam is focused tightest. We can numerically evaluate the $\ee^{-2}$ radius, $\rho_0$, as well as the fraction of total intensity contained within. The total intensity can be evaluated symbolically using the beam's weight function (see Appendix~\ref{app:identities}), and is given by
\begin{widetext}
\begin{equation}
  I_0=\int\modu[2]{\psi}\du[2]{r}=\frac{2\pi{(kb+2)}^2}{k^4k^{2kb}}\ma\int_0^k q^{2kb+1}\du{q}=\frac{2\pi {(kb+2)}^2}{k^2(2kb+2)}\ma.
\end{equation}
\end{widetext}
Figure~\ref{fig:kb_intensity} gives plots for the spot radius, and the fraction of intensity contained. While the zero radius is monotonically decreasing with decreasing $kb$, below $kb\approx 1/2$ the fraction of total intensity drops sharply. In this regime, the initial spot size is a poor focal measure; the beam is highly oscillatory, and a significant proportion of the intensity is contained in the outer rings. As an example, Figure~\ref{fig:b0_2-beam} gives the focal plane beam profile of a beam with $kb=-0.8$, for which only $\approx 33.4\%$ of the intensity is contained in the initial spot.


\section{Defining focal concentration}\label{sec:focal-concentration}
Using second moments to measure beam width is in effect \emph{enhancing} intensity further from the focus. This leads to the aforementioned convergence problems. A different approach is to \emph{penalise} intensity farther from the focus in the denominator of some measure. A common, simple approach is to use the ratio of total intensity to the ratio of intensity in some truncated area. This gives a dimensionless measure of \emph{focal concentration}, which can be converted into an area by multiplying by a consistent choice of area. A sensible choice is the area of truncation itself: if \emph{all} the intensity is located within this area, clearly this area contains the entire focal region. Otherwise, some percentage of the intensity is \emph{not} contained within, and the dimensionless ratio will be greater than 1, thus the focal area will be somewhat larger than the area of truncation.

This can be formulated using distributions: specifically, given an indicator distribution $\mathds{1}_U$ for some truncated region $U$, this measure of focal extent consists of the ratio
\begin{equation}
  \alpha_U=A_U\times\frac{\int \modu[2]{\psi}\du[2]{r}}{\int \mathds{1}_U\modu[2]{\psi}\du[2]{r}},\quad A_U=\int \mathds{1}_U \du[2]{r}.
\end{equation}
For example, if one chooses to truncate to the circle of radius $s$, the indicator distribution is $\Pi(\rho/s)$, where
\begin{equation}\label{eq:pi-step}
  \Pi(x)=\begin{cases}
  1,\quad x\leq 1,
  \\
  0,\quad x>1,
  \end{cases}
\end{equation}
and the area of this indicator is $\pi s^2$. 

Rather than indicator distributions, which are discontinuous, we could chose a continuous distribution. Thus, I propose the following candidate focal area measure, which uses a Gaussian distribution,
\begin{equation}\label{eq:gaussian-focal-conc}
  \alpha(s)=\frac{\pi s^2\int \modu[2]{\psi}\du[2]{r}}{\int \modu[2]{\psi}\ee^{-\rho^2/s^2}\du[2]{r}}.
\end{equation}
I will refer to this measure simply as \emph{focal concentration}.

For the Gaussian beam, the focal concentration is able to be evaluated in closed form, and is given by
\begin{equation}
  \alpha_G(s)=\pi(s^2+\frac{b^2+z^2}{kb})=\pi s^2 +\pi {W_G}^2.
\end{equation}
It is clear that in the limit $s\to 0$ this gives the beam area given by the usual spot size. This limit $s\to 0$ is a particularly useful limit, as in this limit the Gaussian distribution becomes the cylindrical Dirac delta,
\begin{equation}
  \lim_{s\to 0} \frac{1}{\pi s^2}\ee^{-\rho^2/s^2}=\frac{1}{2\pi \rho}\delta(\rho).
\end{equation}
Note that the distribution from~\eqref{eq:pi-step} also reduces to a Dirac Delta,
\begin{equation}
  \lim_{s\to0}\frac{\Pi(\rho/s)}{\pi s^2 }=\frac{1}{2\pi\rho}\delta(\rho).
\end{equation}

The focal area in this limit then becomes the ratio of total intensity to the intensity along the beam axis,
\begin{equation}
  \alpha(0)=\frac{\int \modu[2]{\psi}\du[2]{r}}{\modu[2]{\psi}\big|_{\rho=0}}.
\end{equation}
I will refer to this as the \emph{peak ratio area}. The peak ratio area, as well as focal concentration area, are definable for an arbitrary peak, located at radial coordinate $\vv{\rho_i}$, by simply translating the Gaussian factor (or other appropriate distribution),
\begin{equation}
  \exp\left(-\frac{\rho^2}{s^2}\right)\to\exp\left(-\frac{\modu[2]{\vv{\rho}-\vv{\rho}_i}}{s^2}\right).
\end{equation}
In practice, these measures are most useful for the global maximum of a beam profile.

Closed form expressions for $\alpha(s)$ for the Laguerre-Gauss beams are achievable using a hypergeometric function, and are given by
\begin{widetext}
\begin{equation}
  \alpha_{p,m}(s)=\alpha_G(s)\frac{{\left(s/W_G\right)}^{4|m|}p!  (p+|m|)! {\left({W_G}^2/s^2+1\right)}^{2p+|m|}}{(2p+|m|)! \, _2F_1\left(-p,-p;-2p-|m|;1-s^4/{W_G}^4\right)}.
\end{equation}
\end{widetext}
Here, $W_G$ is the usual spot size of the embedded Gaussian beam. For a derivation see Appendix~\ref{app:focal-concentration}. Note that this expression is significantly more complex compared to the second moment beam width, a drawback of this method. A further limitation is that there can exist beams with different focal concentrations for finite $s$ that nevertheless have identical $s=0$ limits. In particular, this is the case for \emph{all} Laguerre-Gauss beams with $m=0$. All such modes have the same ratio of total intensity to peak intensity, and as such have $\pi {W_G}^2$ as their peak ratio area. 

This limitation is not unique to the peak ratio area, and can in principle apply to the second moment as well. There exist well known counterexamples in probability theory of different probability distributions that have identical moments of all order. For example, the parametrised family of distributions
\begin{equation}
  p_\alpha (x)=\frac{1}{24}\exp(-x^{1/4})(1-\alpha \sin(x^{1/4})),
\end{equation}
have identical moments for $\alpha\in[0,1]$, yet are clearly different functions~\cite[p. 227]{Fellerintroductionprobabilitytheory1971}.

\section{Highest peak for a given intensity}\label{sec:lowest-limit}
I will now derive the highest peak for a given intensity, i.e.~the minimum of $\alpha(0)$, for exact solutions. The limit will be derived using the Cauchy-Schwartz inequality, which will also show us that this limit is unique. Expanding a wave-function using~\eqref{eq:expansion} allows us to express the total intensity, as well as the peak value, in terms of integrals over the beams weight functions (see Appendix~\ref{app:identities}). This gives
\begin{equation}
  \alpha(0) = 2\pi\sum_{m\in\mathbb{Z}}\frac{\int_0^k \modu[2]{f_m(\kappa/k)}\kappa\du{\kappa}}{\modu[2]{\int_0^k f_0(\kappa/k)\kappa\du{\kappa}}}.
\end{equation}
The Cauchy-Schwartz inequality is
\begin{equation}\label{eq:cauchy-schwartz}
  \modu[2]{\int_a^b F(x){G(x)}^* \du{x} } \leq \int_a^b \modu[2]{F(x)}\du{x}\int_a^b \modu[2]{G(x)}\du{x},
\end{equation}
with equality iff.~$F$ and $G$ are scalar multiples. If we set $F(\kappa)=f_0(\kappa/k)\sqrt{\kappa}$, and $G=\sqrt{\kappa}$ in~\eqref{eq:cauchy-schwartz}, we have
\begin{widetext}
\begin{equation}
  \modu[2]{\int_0^k f_0(\kappa/k)\kappa\du{\kappa}}\leq \int_0^k \kappa\du{\kappa}\int_0^k \modu[2]{f_0(\kappa/k)} \kappa\du{\kappa}
  =\frac{k^2}{2}\int_0^k \modu[2]{f_0(\kappa/k)} \kappa\du{\kappa},
\end{equation}
\end{widetext}
with equality iff.~$f(\kappa)$ is constant. We can also note the simple inequality
\begin{equation}
  \frac{k^2}{2}\int_0^k \modu[2]{f_0(\kappa)} \kappa\du{\kappa}\leq\frac{k^2}{2}\sum_{m\in\mathbb{Z}} \int_0^k\modu[2]{f_m(\kappa)}\kappa\du{\kappa},
\end{equation}
with equality iff.~the beam is azimuthally symmetric (i.e.~$f_0$ is the only non-zero component). This gives the final inequality for $\alpha$,
\begin{equation}
  \alpha(0)\geq \frac{4\pi}{k^2},
\end{equation}
with equality iff.~the wave is azimuthally symmetric, with $f(\kappa/k)$ constant, evaluated in the focal plane $z=0$. Such a beam is referred to as the `proto-beam' by Lekner, who gives an extensive study of its properties~\cite{LeknerTightfocusinglight2016}. This beam is the confluent limit $b\to0$ of two families of beams, those with exponential weight functions,
\begin{equation}
  f(b,\kappa/k)\propto\ee^{q b},\quad q=\sqrt{k^2-\kappa^2}
\end{equation}
and those with Gaussian weight functions,
\begin{equation}
  f(b,\kappa/k)\propto \ee^{-b\kappa^2/2k}.
\end{equation}
These families approximate the Gaussian beam along the beam axis, and radially, respectively, for large $kb$~\cite{Andrejic2017,Lekner2018}. The limit of tightest focus for both of these families is the proto-beam.

I have now shown that the proto-beam is the \emph{unique} single peaked scalar beam that achieves the smallest peak ratio area, and thus, it is the unique beam that for a given total intensity, achieves the highest possible peak intensity.

\begin{figure}[hbt]
  \includegraphics[width=5in]{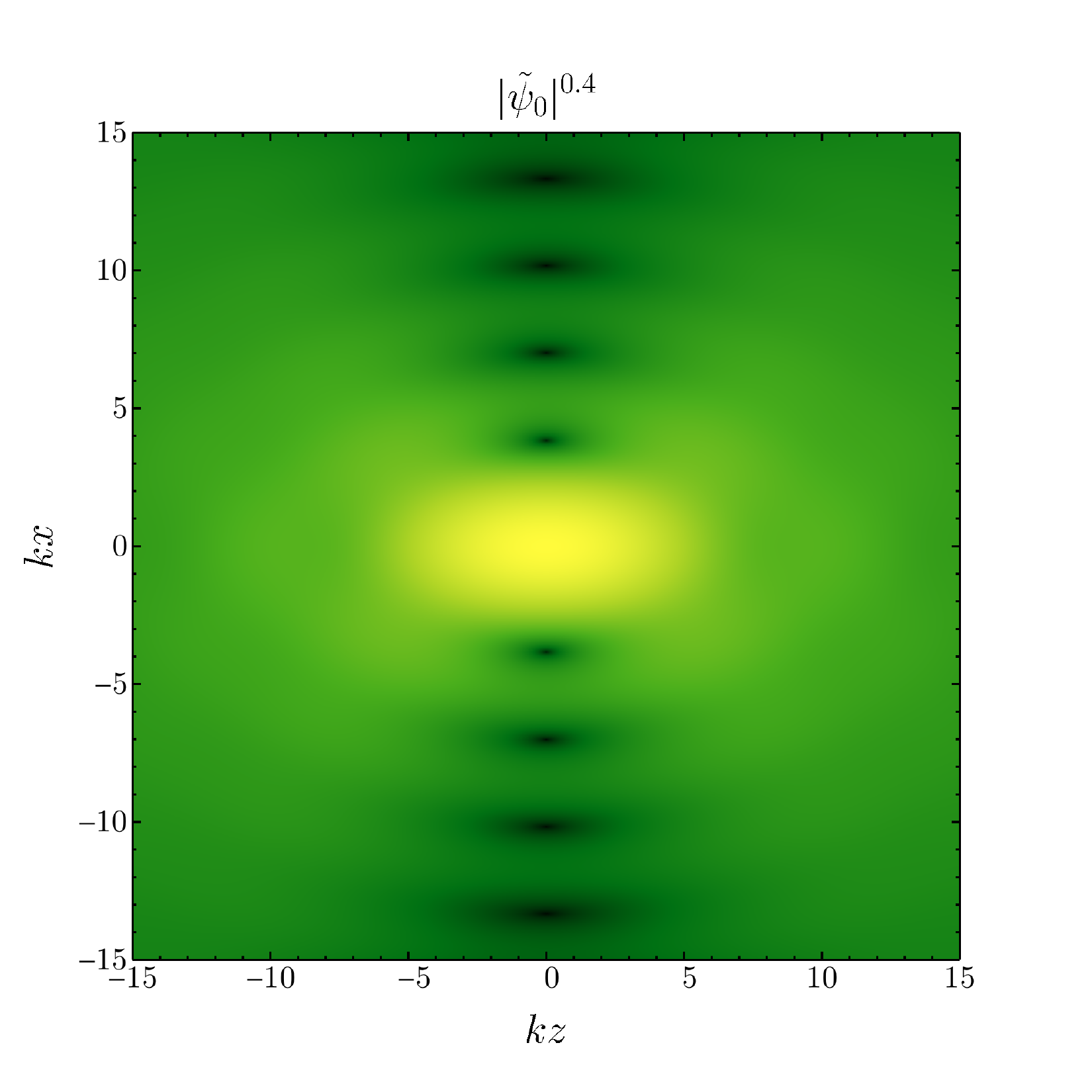}
  \caption{Proto-beam modulus, scaled to the power of $0.4$. Lighter shading indicates larger modulus. Note the black zeros in the focal plane $z=0$.}\label{fig:proto}
\end{figure}

For posterity, Figure~\ref{fig:proto} gives the radial profile of the proto-beam in the focal plane, proportional to a Jinc function, where $\mathop{\mathrm{Jinc}}(x)= J_1(x)/x$.
Note that unlike the Airy disk pattern, the proto-beam is only proportional to a Jinc function in the \emph{focal plane}, and does not have zeros elsewhere.
This does not imply that the proto-beam has the smallest peak ratio area \emph{everywhere}. There is a trade-off to be made with divergence angle and focal extent; generally reducing focal extent also requires a larger divergence angle. As we shall see in Section~\ref{sec:divergence-angles}, the proto-beam has a comparatively large divergence angle of $45^\circ$.  

\section{Divergence angles for the proto-beam}\label{sec:divergence-angles}
\begin{figure}[hbt]
  \centering
  \includegraphics[width=7in]{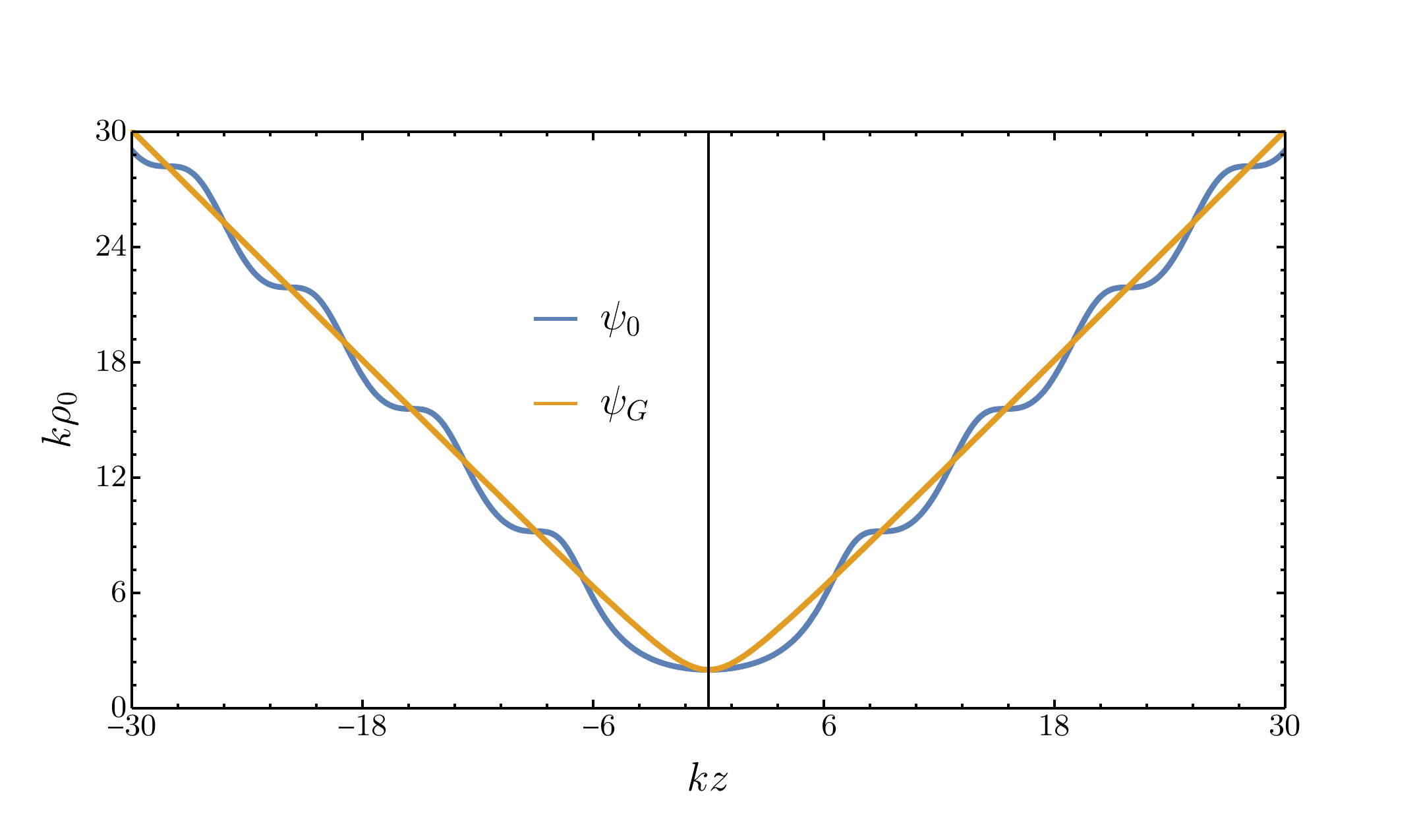}
  \caption{Comparison of spot size dependence on $z$ for proto-beam, and the $kb=2$ Gaussian. Note how the average spot size of the proto-beam is given by the Gaussian spot size, and the divergence angle is $45^\circ$. Interestingly, the Gaussian's minimum spot size of $k\rho_0=2$ is the same as that of the proto-beam.}\label{fig:comparison}
\end{figure}

I will now demonstrate how one may use the change of $\alpha(0)$ with $z$ to compute the divergence angle for the proto-beam. For the proto-beam, the axial wave-function is given by
\begin{equation}
  \tpsi_0(0,z)=\frac{2}{k^2}\int_0^k\ee^{\ii q z} q\du{q}=\frac{2}{k^2 z^2}\left(\ee^{\ii kz}(1-\ii k z)-1\right).
\end{equation}
The modulus is given by
\begin{equation}
  \frac{8}{k^4z^4}\left(1+\frac{k^2z^2}{2}-\cos(kz)-kz\sin(kz)\right).
\end{equation}
The total intensity is given by $4\pi/k^2$. If we interpret the area parameter as the area of a spot with radius $\rho_0$, i.e.~$\alpha = \pi {\rho_0}^2$, the spot radius is given by
\begin{equation}
  \rho_0(z)=\frac{kz^2}{\sqrt{2}}{\left(1+\frac{k^2z^2}{2}-\cos(kz)-kz\sin(kz)\right)}^{-1/2}
\end{equation}
Compare this with the Gaussian beam, $W_G=\sqrt{2(b^2+z^2)/kb}$, and its divergence ratio of $\sqrt{2/kb}$. Figure~\ref{fig:comparison} gives a plot of the spot size as a function of $z$, compared with a Gaussian beam with a $45^\circ$ ($kb=2$) divergence angle. 

A visual inspection suggests that for sufficiently large $kz$, the proto-beam spot size can be approximated as a linear function plus an oscillatory term,
\begin{equation}
  \rho_0(z)\approx f(z)=a_0 + a_1 kz + b_0\cos(kz+b_1),\quad kz\gg 1.
\end{equation}
For a function of this form, the limit of $f(z)/z$ as $z\to\infty$ gives $a_1$, with the oscillations decaying to zero, and hence we can find the average slope $m$ of $\rho_0$ via
\begin{widetext}
\begin{equation}
  m^2=\lim_{z\to\infty}\frac{\rho_0{(z)}^2}{z^2}=\lim_{z\to\infty}\frac{k^2z^2}{2}{\left(1+\frac{k^2z^2}{2}-\cos(kz)-kz\sin(kz)\right)}^{-1}.
\end{equation}
\end{widetext}
Substituting the maximum and minimum values of $\cos$ and $\sin$, we get the following inequality,
\begin{equation}
  \lim_{z\to\infty}\frac{k^2z^2}{2+k^2z^2-2-2kz}\geq m^2\geq\lim_{z\to\infty}\frac{k^2z^2}{2+k^2z^2+2+2kz}.
\end{equation}
The sandwich theorem gives us $m=1$,
\begin{equation}
  \lim_{z\to\infty}\frac{k^2}{k^2-2k/z}=1\geq m^2 \geq \lim_{z\to\infty} \frac{k^2}{k^2-2k/z-4/z^2}=1.
\end{equation}
Therefore, the average slope of $\rho_0(z)$ is 1, which corresponds to a divergence angle of $45^\circ$. It is interesting to note that Lekner has also studied pulses formed by coherent superpositions of proto-beam wavefunctions of different frequencies~\cite{Lekner2018}. For these pulses he has shown that the divergence angle of the pulse is asymptotically $45^\circ$, which is consistent with our result for the proto-beam.

\section{Comparison of Porras beam to proto-beam}\label{sec:porras-to-proto-comparison}
\begin{figure}[hbt]
\centering
\subfigure[\label{sfig:step-reg}]{\includegraphics[width=3.2in]{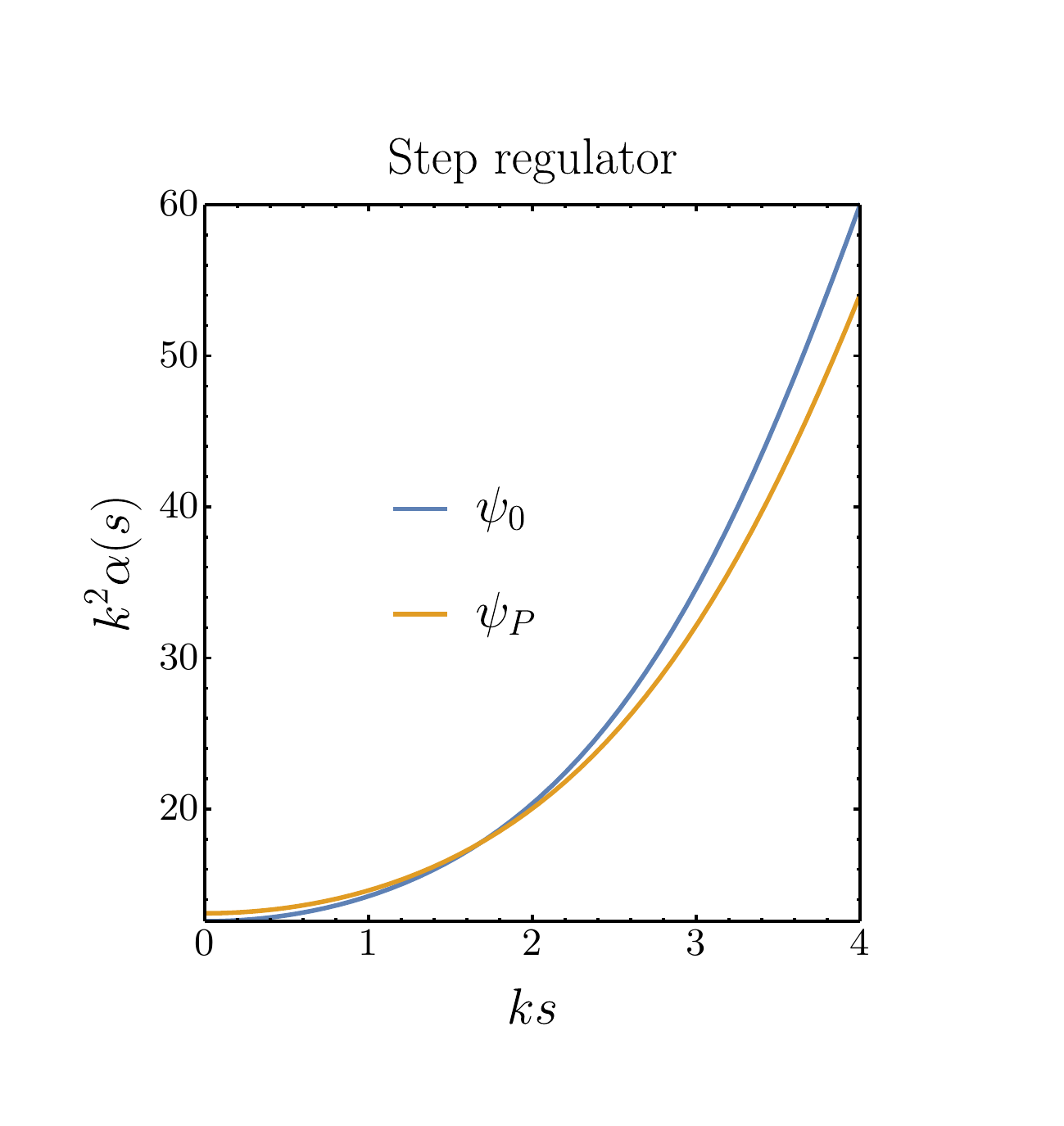}}
\subfigure[\label{sfig:step-delta}]{\includegraphics[width=3.2in]{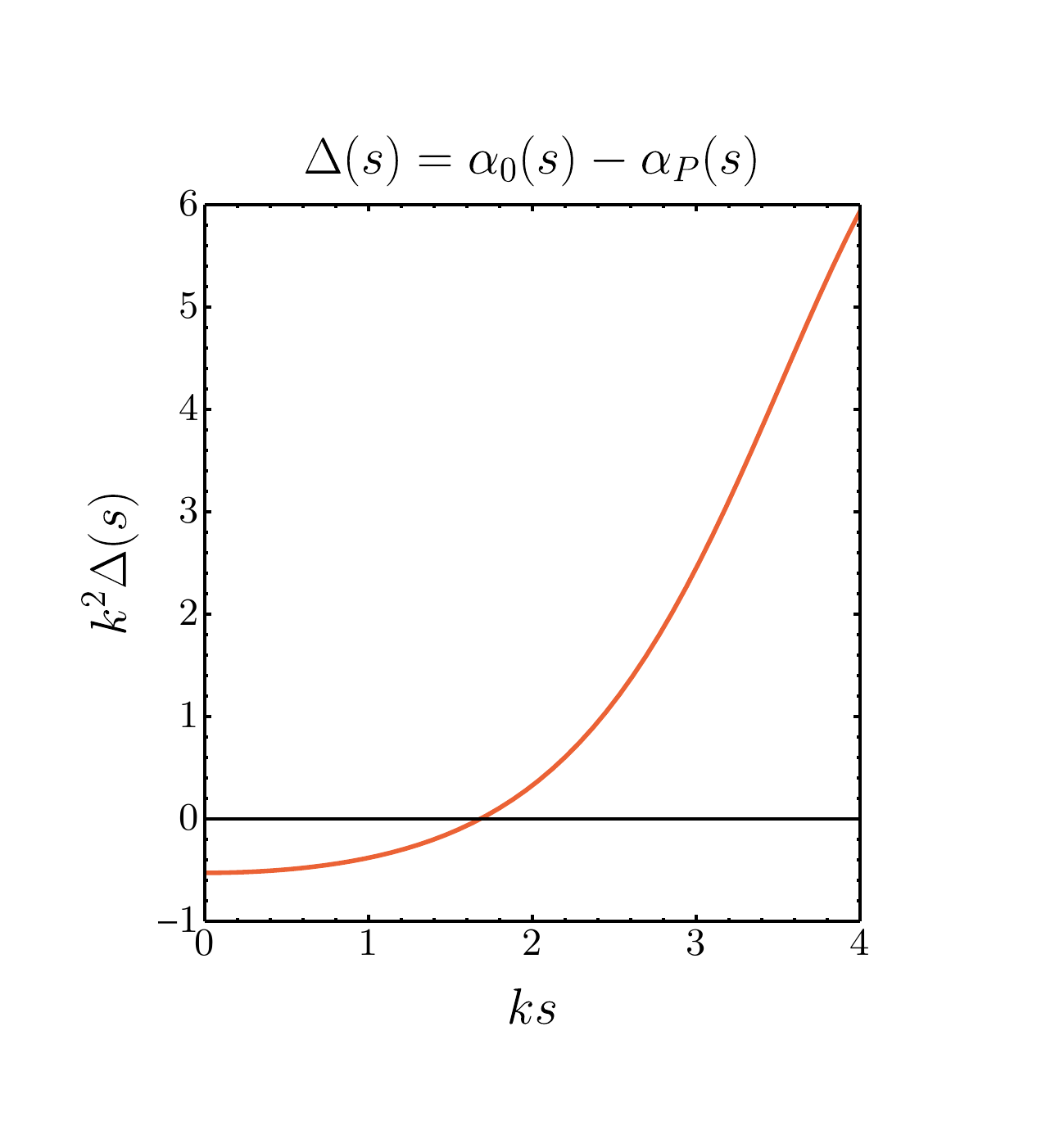}}
  \caption{Step function regulated focal area for the proto beam $\psi_0$, and Porras beam $\psi_P$,~\subref{sfig:step-reg}, and difference in areas,~\subref{sfig:step-delta}. At $ks\approx1.68$ the Porras beam becomes more tightly focused by this measure. This illustrates how using a step regulator for measuring focal area is ambiguous as where one chooses the cut-off to be can determine which beam is more tightly focused.}\label{fig:step-reg}
\end{figure}

\begin{figure}[hbt]
\centering
\includegraphics[width=4in]{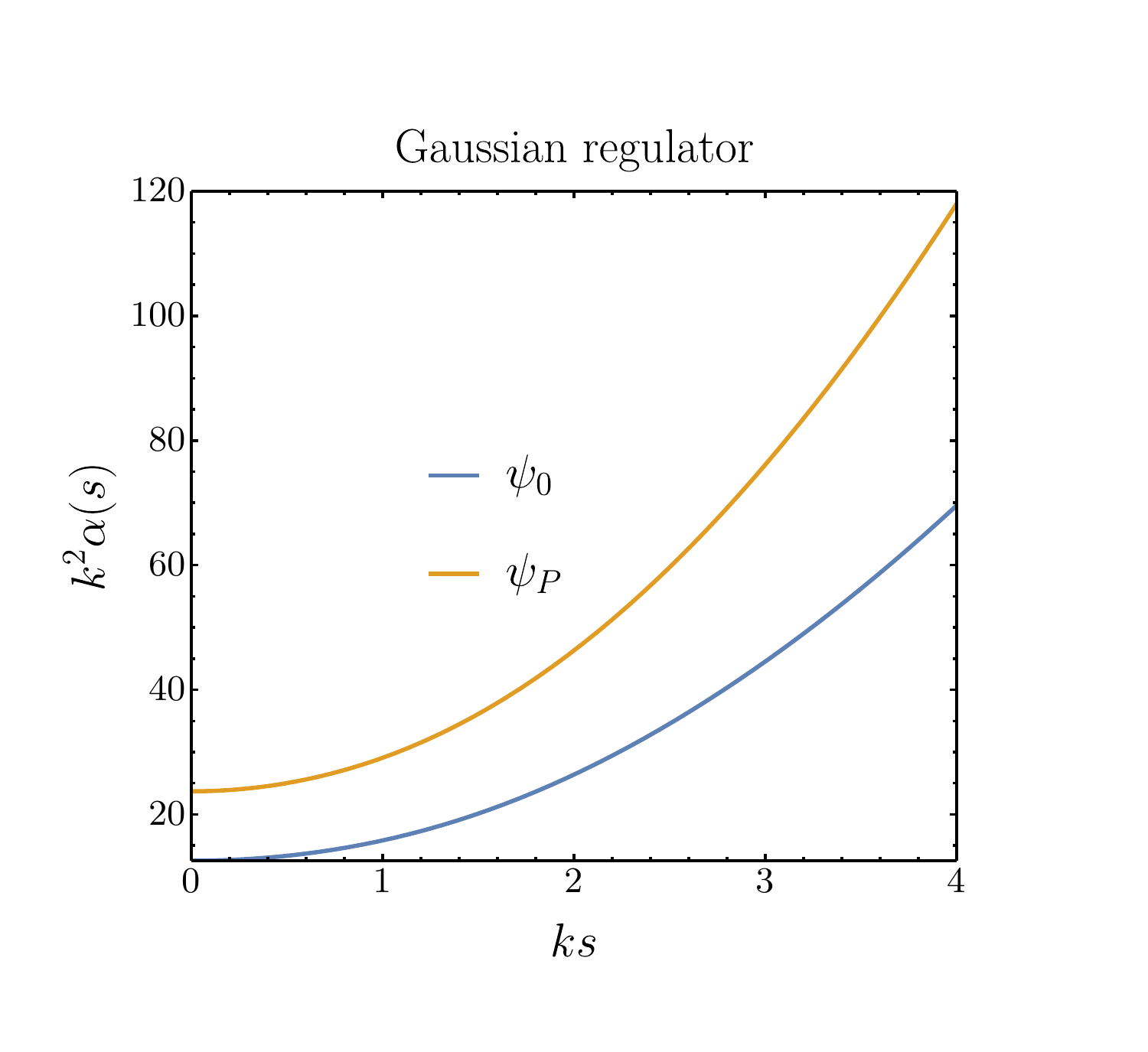}
\caption{Gaussian regulated focal area for the proto beam $\psi_0$, and Porras beam $\psi_P$,~\subref{sfig:step-reg}, and difference in areas,~\subref{sfig:step-delta}. With this choice of regulator, the proto-beam is more tightly focused even up to a regulator length of $ks=4$.}\label{fig:gauss-reg}
\end{figure}

Let us now compare the proto-beam to the Porras beam to demonstrate the ambiguities that arise in comparing different focal measures. For a superposition of Bessel $J_0$ beams, we can express the regulated focal concentration in terms of the weight functions,
\begin{widetext}
\begin{equation}
  \alpha(s)=\frac{\pi s^2\int_{0}^{k}\kappa \du{\kappa}\modu[2]{f(\kappa/k)}}{\iint_{0}^{k}\kappa_1\kappa_2\du{\kappa_1}\du{\kappa_2}f(\kappa_1)\bar{f}(\kappa_2)\Theta(s,\kappa_1,\kappa_2)}.
\end{equation}
\end{widetext}
Here, $\Theta$ is a function decided by the choice of regulator $R(s,\rho)$, that is defined via
\begin{equation}
  \Theta(s,\kappa_1,\kappa_2)\equiv \int_0^\infty\rho\du{\rho}J_0(\kappa_1\rho)J_0(\kappa_2\rho)R(s,\rho).
\end{equation}
For the Gaussian regulator in~\eqref{eq:gaussian-focal-conc}, an application of Weber's second exponential integral~\cite[p.~395]{WatsonTreatise1922} gives
\begin{equation}
  \Theta_G(s,\kappa_1,\kappa_2)=\frac{s^2}{2}\exp\left(-\frac{s^2(\kappa_1^2+\kappa_2^2)}{4}\right)I_0\left(\frac{s^2\kappa_1\kappa_2}{2}\right).
\end{equation}
For the step regulator from~\eqref{eq:pi-step} we get
\begin{widetext}
\begin{equation}
  \Theta_0(s,\kappa_1,\kappa_2)=\frac{s^2}{\kappa_1^2-\kappa_2^2}\exp\left(-\frac{s^2(\kappa_1^2+\kappa_2^2)}{4}\right)\left(\kappa_1^2J_0(\kappa_2 s)\frac{J_1(\kappa_1 s)}{\kappa_1 s}-\kappa_2^2J_0(\kappa_1 s)\frac{J_1(\kappa_2 s)}{\kappa_2 s}\right).
\end{equation}
\end{widetext}
These allow the numeric evaluation for a given beam. In Figures~\ref{fig:step-reg} and~\ref{fig:gauss-reg} we see the dependence of $\alpha$ with $s$ for the proto-beam compared with the Porras beam~\eqref{eq:porras-beam}, for both the Gaussian regulator and step regulators. Both regulators give the same limit at $s=0$, and we can see that the focal area of the proto-beam is smaller at this limit. For the step regulator, there is a crossing point, a value of $ks$ where the focal area of the Porras beam becomes smaller than the proto-beam. On the other hand, choosing a Gaussian regulator, there is no such crossing, at least for $ks\leq 4$. This demonstrates that the relative focal extent of beams is strongly dependent on the choice of measure used.

\section{Limit of peak energy density for electromagnetic beams}\label{sec:electromagnetic}
For electromagnetic beams, the intensity is measured by the energy content of the beam. I will attempt to derive a lower limit for peak energy density for a given energy per unit length of a beam. However, the situation is complicated by the vectorial nature of the beam. In Coulomb gauge, we have $\nabla\cdot \vv{A}=0$, and we can expand the electromagnetic vector potential into a superposition over two orthogonal polarisation modes,
\begin{equation}
	\vv{A}(\vv{r},t)=A_0\sum_{\lambda=1,2}\int\du{\Omega}f_\lambda (\alpha,\beta)\hat{e}_\lambda(\alpha,\beta)\ee^{\ii (\vv{k}\cdot\vv{r}-k ct)},
\end{equation}
where $\int\du{\Omega}=\int_0^{\pi}\int_0^{2\pi}\sin\alpha\du{\alpha}\du{\beta}$ is an integral over solid angles. Here, $A_0$ is to be interpreted as having the dimensionality of electromagnetic potential. For free space propagation we can always fix the gauge so that the scalar potential is zero. A particular choice of orthonormal basis is orthonormal linear polarisation,
\begin{equation}
\begin{aligned}
	\hat{e}_1&=\sin\beta \, \hat{x}-\cos\beta \, \hat{y},
	\\
	\hat{e}_2&=\cos\alpha\cos\beta \, \hat{x}+\cos\alpha\sin\beta \, \hat{y}-\sin\alpha \, \hat{z},
	\\
	\hat{k}&=\sin\alpha\cos\beta \, \hat{x}+\sin\alpha\sin\beta \, \hat{y}+\cos\alpha\hat{z}.
\end{aligned}
\end{equation}
The (complex) electric and magnetic fields are then given by
\begin{equation}
  \begin{aligned}
	&\vv{E}(\vv{r},t) = -\pdv{\vv{A}}{t}=\ii k\vv{A}(\vv{r},t),
	\\
    & \vv{B}(\vv{r},t) = \ii kA_0\sum_{\lambda=1,2}\int\du{\Omega}
	  f_\lambda (\alpha,\beta)
	  \hat{k}\times\hat{e}_\lambda
	  \ee^{\ii (\vv{k}\cdot\vv{r}-kc t)}.
  \end{aligned}
\end{equation}
It is worth noting that $\hat{k}\times\hat{e}_1=\hat{e}_2$, and $\hat{k}\times\hat{e}_2=-\hat{e}_1$. For forward propagating beams, the range of $\alpha$ is restricted to $[0,\pi/2]$. The cycle average energy per unit length of the beam is given by
\begin{equation}
	U=\frac{1}{8\pi}\int(\modu[2]{E}+\modu[2]{B})\du[2]{r},
\end{equation}
The plane wave orthogonality relation in Appendix~\ref{app:radial-integral} allows us to calculate the energy per unit length in terms of the weight functions,
\begin{equation}
	U=\pi\ma\sum_{\lambda=1,2}\int\du{\Omega}\frac{\modu[2]{f_\lambda(\alpha,\beta)}}{\cos\alpha}.
\end{equation}
We can derive a similar inequality as in Section~\ref{sec:focal-concentration}, by noting that the Cauchy-Schwartz inequality can be applied to complex vector fields $\vv{F}$, $\vv{G}$ with a measure $\du{\mu}$ via
\begin{equation}\label{eq:vector-cauchy}
	\modu[2]{\int\vv{F}\cdot\vv{G}^*}\du{\mu}\leq \int\modu[2]{F}\du{\mu}\int\modu[2]{G}\du{\mu}.
\end{equation}
Here, we have equality iff.~$\vv{F}$ and $\vv{G}$ are scalar multiples, as usual. We can apply this to electromagnetic beams by noting that the field at the origin is an integral of a vector field over a sphere. Specifically, the term $\sum_{\lambda=1,2} f_\lambda(\alpha,\beta)\hat{e}_\lambda(\alpha,\beta)$ is to be considered a complex valued, three dimensional vector field over the surface parametrised by $\alpha,\beta$ (namely, the unit sphere), with the integration measure being the spherical measure $\sin\alpha\du{\alpha}\du{\beta}$. We can choose $\vv{F}$ and $\vv{G}$ in~\eqref{eq:vector-cauchy} to be 
\begin{equation}
	\vv{F}=\frac{kA_0}{\sqrt{\cos\alpha}}\sum_{\lambda=1,2}f_\lambda(\alpha,\beta)\hat{e}_\lambda(\alpha,\beta),
	\quad \vv{G}=\sqrt{\cos\alpha}\, \hat{n}.
\end{equation}
Here, $\hat{n}$ is a constant unit vector field that is in the same direction as $\vv{E}$ at the origin. This gives
\begin{equation}
	k^2\ma\modu[2]{\sum_{\lambda=1,2}\int\du{\Omega}f_\lambda \hat{e}_\lambda}
  \leq 
  \pi k^2\ma\sum_{\lambda=1,2}
  \int\du{\Omega}\frac{\modu[2]{f_\lambda}}{\cos\alpha}.
\end{equation}
Note that there are no terms proportional to the product of $f_1,f_2$ as their polarisation vectors are orthogonal when evaluated at the same angle $\alpha,\beta$. The LHS is equal to the modulus square of the electric field at the origin, $\modu[2]{E_0}$, and the RHS is $k^2 U$. An identical inequality can be found for $\vv{B}$, and the combination of these inequalities gives us
\begin{equation}
	\modu[2]{E_0}+\modu[2]{B_0}\leq2k^2 U.
\end{equation}
This can be rewritten as
\begin{equation}\label{eq:electromagnetic-parameter}
	8\pi  u_0\leq 2k^2 U ,\quad \frac{U}{u_0}\geq \frac{4\pi}{k^2}.
\end{equation}
Thus, we have obtained the exact same inequality as the scalar case. Unlike the scalar case, equality cannot be reached for a physical electromagnetic beam. For this to occur, $\vv{\Psi}(\alpha,\beta)=\sum_\lambda f_\lambda(\alpha,\beta) \hat{e}_\lambda(\alpha,\beta)$ would have to be a constant direction, constant magnitude vector field. This defines a system of partial differential equations,
\begin{equation}
	\pdv{\vv{\Psi}}{\alpha}=0,\quad \pdv{\vv{\Psi}}{\beta}=0.
\end{equation}
This system however, can be shown to be inconsistent, and the only solution is the trivial solution $f_1=f_2=0$. As the equality in~\eqref{eq:electromagnetic-parameter} is not reachable, it remains unclear what the true limit of peak ratio area is for electromagnetic beams. It is quite possible that a different lower limit exists for different forms of vector potential, e.g.~linear compared with circular polarised vector potentials.

\section{Discussion}\label{sec:discussion}
Existing measures of focal extent remain adequate for paraxial beams. In particular, the relation of the second moment measure to the divergence angle, and beam equality for paraxial beams is a strong physical justification for its use. However, in the paraxial case there is not lower limit of focal extent to consider, as paraxial solutions can be arbitrarily tightly focused. For exact solutions, eventually the diffraction limit is reached. 

There are a few related, but different bounds to place on the focal properties of a beam. One of these is the highest peak achievable for a given total intensity. The peak ratio area discussed in Section~\ref{sec:lowest-limit} shows that the highest peak achievable is that of the proto-beam. As discussed previously, Porras has placed the lowest bound on second moment width, and the beam that achieves this is distinct from the proto-beam. As the proto-beam has a divergent second moment, one might assume it is less tightly focused than the Porras beam. However, as we have seen in Section~\ref{sec:porras-to-proto-comparison}, which beam is to be considered more tightly focused depends strongly on the choice of focal measure. 

Choosing the Gaussian regulator as a measure of focal intensity, I was not able to find a crossing point as in Figure~\ref{sfig:step-delta}, up to $ks=4$. However, this does not discount the possibility that some crossing may occur. It is worth noting however, that the area defined using such a regulator is an absolutely monotonic function of $s$, and the derivative of $\alpha_P(s)$ is greater than $\alpha_0(s)$ over the region $0\leq ks\leq4$. It is possible this may suffice to show that there is no crossing, however I am not aware of a relevant bound to apply.

I have largely restricted my attention to scalar beams. There is already a sizeable body of work on the tight focusing of electromagnetic beams, such as the work of Dorn, Quabis \emph{et al}~\cite{DornSharperFocusRadially2003,QuabisFocusinglighttighter2000a,Quabisfocuslighttheoretical2001}, Kozawa and Sato~\cite{KozawaSharperfocalspot2007}, and Arkhipov and Schulten~\cite{ArkhipovLimitsreductioneffective2009}. In particular Dorn, Quabis \emph{et al} and Kozawa and Sato use full width half maximum as a spot size for electromagnetic beams, with all the benefits and drawbacks mentioned in~\ref{sec:introduction}. Arkhipov and Schulten are concerned with the focal volume, defined via point spread functions. In my view, given that full width half maximum is inappropriate for highly oscillatory beams, the second moment is more appropriate. An extension of the variational technique of Porras~\cite{Porrasbestqualityoptical1994} could potentially be used to find the electromagnetic beam with the smallest second moment.

In Section~\ref{sec:electromagnetic} I attempted to derive a bound on peak energy density for electromagnetic beams. However, as I have noted in said section, this lower bound is not reached, and the true lower bound is somewhat higher. Further study is required to refine this bound. It is likely that the bound is dependent on the polarisation of the beam. For example, Dorn \emph{et al} investigated the spot size of electromagnetic beams passing through annular apertures, with their definition of spot size being the area contained in the half of peak intensity contour. They found a spot size of approximately $0.17\lambda^2$ for radial, $0.26\lambda^2$ for linear, and $0.22\lambda^2$ for circular polarisations. Although the type of focal measure being investigated is very different, I expect a similar result of a different maximum peak intensity for a given total, for different polarisations. However, as we have seen in the scalar case, hollow modes have a lower peak intensity compared with azimuthally symmetric modes (see Section~\ref{sec:lowest-limit}). I expect this will also be the case for the electromagnetic beams.

\begin{acknowledgments}
The author would like to thank EProf. John Lekner for his helpful feedback.
\end{acknowledgments}
\bibliography{refs,iso}

\begin{thebibliography}{24}%
\makeatletter
\providecommand \@ifxundefined [1]{%
 \@ifx{#1\undefined}
}%
\providecommand \@ifnum [1]{%
 \ifnum #1\expandafter \@firstoftwo
 \else \expandafter \@secondoftwo
 \fi
}%
\providecommand \@ifx [1]{%
 \ifx #1\expandafter \@firstoftwo
 \else \expandafter \@secondoftwo
 \fi
}%
\providecommand \natexlab [1]{#1}%
\providecommand \enquote  [1]{``#1''}%
\providecommand \bibnamefont  [1]{#1}%
\providecommand \bibfnamefont [1]{#1}%
\providecommand \citenamefont [1]{#1}%
\providecommand \href@noop [0]{\@secondoftwo}%
\providecommand \href [0]{\begingroup \@sanitize@url \@href}%
\providecommand \@href[1]{\@@startlink{#1}\@@href}%
\providecommand \@@href[1]{\endgroup#1\@@endlink}%
\providecommand \@sanitize@url [0]{\catcode `\\12\catcode `\$12\catcode
  `\&12\catcode `\#12\catcode `\^12\catcode `\_12\catcode `\%12\relax}%
\providecommand \@@startlink[1]{}%
\providecommand \@@endlink[0]{}%
\providecommand \url  [0]{\begingroup\@sanitize@url \@url }%
\providecommand \@url [1]{\endgroup\@href {#1}{\urlprefix }}%
\providecommand \urlprefix  [0]{URL }%
\providecommand \Eprint [0]{\href }%
\providecommand \doibase [0]{http://dx.doi.org/}%
\providecommand \selectlanguage [0]{\@gobble}%
\providecommand \bibinfo  [0]{\@secondoftwo}%
\providecommand \bibfield  [0]{\@secondoftwo}%
\providecommand \translation [1]{[#1]}%
\providecommand \BibitemOpen [0]{}%
\providecommand \bibitemStop [0]{}%
\providecommand \bibitemNoStop [0]{.\EOS\space}%
\providecommand \EOS [0]{\spacefactor3000\relax}%
\providecommand \BibitemShut  [1]{\csname bibitem#1\endcsname}%
\let\auto@bib@innerbib\@empty
\bibitem [{\citenamefont {Saleh}\ and\ \citenamefont
  {Teich}(2001)}]{SalehFundamentalsphotonics2001}%
  \BibitemOpen
  \bibfield  {author} {\bibinfo {author} {\bibfnamefont {B.~E.~A.}\
  \bibnamefont {Saleh}}\ and\ \bibinfo {author} {\bibfnamefont {M.~C.}\
  \bibnamefont {Teich}},\ }\href@noop {} {\emph {\bibinfo {title} {Fundamentals
  of Photonics}}}\ (\bibinfo  {publisher} {{Wiley-Interscience}},\ \bibinfo
  {address} {Hoboken, NJ},\ \bibinfo {year} {2001})\BibitemShut {NoStop}%
\bibitem [{\citenamefont {Novotny}\ and\ \citenamefont
  {Hecht}(2012)}]{NovotnyPrinciplesnanooptics2012}%
  \BibitemOpen
  \bibfield  {author} {\bibinfo {author} {\bibfnamefont {L.}~\bibnamefont
  {Novotny}}\ and\ \bibinfo {author} {\bibfnamefont {B.}~\bibnamefont
  {Hecht}},\ }\href@noop {} {\emph {\bibinfo {title} {Principles of
  Nano-Optics}}},\ \bibinfo {edition} {2nd}\ ed.\ (\bibinfo  {publisher}
  {{Cambridge University Press}},\ \bibinfo {address} {Cambridge},\ \bibinfo
  {year} {2012})\BibitemShut {NoStop}%
\bibitem [{\citenamefont {Siegman}(1986)}]{SiegmanLasers1986}%
  \BibitemOpen
  \bibfield  {author} {\bibinfo {author} {\bibfnamefont {A.~E.}\ \bibnamefont
  {Siegman}},\ }\href@noop {} {\emph {\bibinfo {title} {Lasers}}}\ (\bibinfo
  {publisher} {{University Science Books}},\ \bibinfo {address} {California},\
  \bibinfo {year} {1986})\BibitemShut {NoStop}%
\bibitem [{ISO 11146:2005(2005)}]{iso11146}%
  \BibitemOpen
  ISO 11146:2005,\ \href@noop {} {\enquote {\bibinfo {title} {Lasers and
  laser-related equipment — test methods for laser beam widths, divergence
  angles and beam propagation ratios},}\ } (\bibinfo {year} {2005})\BibitemShut
  {NoStop}%
\bibitem [{\citenamefont {Phillips}\ and\ \citenamefont
  {Andrews}(1983)}]{PhillipsSpotsizedivergence1983}%
  \BibitemOpen
  \bibfield  {author} {\bibinfo {author} {\bibfnamefont {R.~L.}\ \bibnamefont
  {Phillips}}\ and\ \bibinfo {author} {\bibfnamefont {L.~C.}\ \bibnamefont
  {Andrews}},\ }\bibfield  {title} {\enquote {\bibinfo {title} {Spot size and
  divergence for {{Laguerre Gaussian}} beams of any order},}\ }\href {\doibase
  10.1364/AO.22.000643} {\bibfield  {journal} {\bibinfo  {journal} {Applied
  Optics}\ }\textbf {\bibinfo {volume} {22}},\ \bibinfo {pages} {643} (\bibinfo
  {year} {1983})}\BibitemShut {NoStop}%
\bibitem [{\citenamefont {Saghafi}\ and\ \citenamefont
  {Sheppard}(1998)}]{Saghafibeampropagationfactor1998}%
  \BibitemOpen
  \bibfield  {author} {\bibinfo {author} {\bibfnamefont {S.}~\bibnamefont
  {Saghafi}}\ and\ \bibinfo {author} {\bibfnamefont {C.~J.~R.}\ \bibnamefont
  {Sheppard}},\ }\bibfield  {title} {\enquote {\bibinfo {title} {The beam
  propagation factor for higher order {{Gaussian}} beams},}\ }\href {\doibase
  10.1016/S0030-4018(98)00256-9} {\bibfield  {journal} {\bibinfo  {journal}
  {Optics Communications}\ }\textbf {\bibinfo {volume} {153}},\ \bibinfo
  {pages} {207--210} (\bibinfo {year} {1998})}\BibitemShut {NoStop}%
\bibitem [{\citenamefont {Lavi}, \citenamefont {Prochaska},\ and\ \citenamefont
  {Keren}(1988)}]{LaviGeneralizedbeamparameters1988}%
  \BibitemOpen
  \bibfield  {author} {\bibinfo {author} {\bibfnamefont {S.}~\bibnamefont
  {Lavi}}, \bibinfo {author} {\bibfnamefont {R.}~\bibnamefont {Prochaska}}, \
  and\ \bibinfo {author} {\bibfnamefont {E.}~\bibnamefont {Keren}},\ }\bibfield
   {title} {\enquote {\bibinfo {title} {Generalized beam parameters and
  transformation laws for partially coherent light},}\ }\href {\doibase
  10.1364/AO.27.003696} {\bibfield  {journal} {\bibinfo  {journal} {Applied
  Optics}\ }\textbf {\bibinfo {volume} {27}},\ \bibinfo {pages} {3696}
  (\bibinfo {year} {1988})}\BibitemShut {NoStop}%
\bibitem [{\citenamefont {Siegman}(1990)}]{SiegmanNewdevelopmentslaser1990}%
  \BibitemOpen
  \bibfield  {author} {\bibinfo {author} {\bibfnamefont {A.~E.}\ \bibnamefont
  {Siegman}},\ }\bibfield  {title} {\enquote {\bibinfo {title} {New
  developments in laser resonators},}\ }in\ \href {\doibase 10.1117/12.18425}
  {\emph {\bibinfo {booktitle} {Optical {{Resonators}}}}},\ Vol.\ \bibinfo
  {volume} {1224},\ \bibinfo {editor} {edited by\ \bibinfo {editor}
  {\bibfnamefont {D.~A.}\ \bibnamefont {Holmes}}}\ (\bibinfo  {publisher}
  {{SPIE}},\ \bibinfo {year} {1990})\ p.~\bibinfo {pages} {2}\BibitemShut
  {NoStop}%
\bibitem [{\citenamefont
  {Siegman}(1993)}]{SiegmanDefiningmeasuringoptimizing1993}%
  \BibitemOpen
  \bibfield  {author} {\bibinfo {author} {\bibfnamefont {A.~E.}\ \bibnamefont
  {Siegman}},\ }\bibfield  {title} {\enquote {\bibinfo {title} {Defining,
  measuring, and optimizing laser beam quality},}\ }in\ \href {\doibase
  10.1117/12.150601} {\emph {\bibinfo {booktitle} {Proceedings of {{SPIE}}}}},\
  Vol.\ \bibinfo {volume} {1868},\ \bibinfo {editor} {edited by\ \bibinfo
  {editor} {\bibfnamefont {A.}~\bibnamefont {Bhowmik}}}\ (\bibinfo  {publisher}
  {{SPIE}},\ \bibinfo {address} {Los Angeles, California},\ \bibinfo {year}
  {1993})\ pp.\ \bibinfo {pages} {2--12}\BibitemShut {NoStop}%
\bibitem [{\citenamefont {Porras}(1994)}]{Porrasbestqualityoptical1994}%
  \BibitemOpen
  \bibfield  {author} {\bibinfo {author} {\bibfnamefont {M.~A.}\ \bibnamefont
  {Porras}},\ }\bibfield  {title} {\enquote {\bibinfo {title} {The best quality
  optical beam beyond the paraxial approximation},}\ }\href {\doibase
  10.1016/0030-4018(94)90475-8} {\bibfield  {journal} {\bibinfo  {journal}
  {Optics Communications}\ }\textbf {\bibinfo {volume} {111}},\ \bibinfo
  {pages} {338--349} (\bibinfo {year} {1994})}\BibitemShut {NoStop}%
\bibitem [{\citenamefont {Levin}(1996)}]{LevinFastintegrationrapidly1996a}%
  \BibitemOpen
  \bibfield  {author} {\bibinfo {author} {\bibfnamefont {D.}~\bibnamefont
  {Levin}},\ }\bibfield  {title} {\enquote {\bibinfo {title} {Fast integration
  of rapidly oscillatory functions},}\ }\href {\doibase
  10.1016/0377-0427(94)00118-9} {\bibfield  {journal} {\bibinfo  {journal}
  {Journal of Computational and Applied Mathematics}\ }\textbf {\bibinfo
  {volume} {67}},\ \bibinfo {pages} {95--101} (\bibinfo {year}
  {1996})}\BibitemShut {NoStop}%
\bibitem [{\citenamefont {Levin}(1982)}]{LevinProcedurescomputingone1982}%
  \BibitemOpen
  \bibfield  {author} {\bibinfo {author} {\bibfnamefont {D.}~\bibnamefont
  {Levin}},\ }\bibfield  {title} {\enquote {\bibinfo {title} {Procedures for
  computing one- and two-dimensional integrals of functions with rapid
  irregular oscillations},}\ }\href {\doibase
  10.1090/S0025-5718-1982-0645668-7} {\bibfield  {journal} {\bibinfo  {journal}
  {Math. Comp.}\ }\textbf {\bibinfo {volume} {38}},\ \bibinfo {pages}
  {531--538} (\bibinfo {year} {1982})}\BibitemShut {NoStop}%
\bibitem [{\citenamefont
  {Feller}(1971)}]{Fellerintroductionprobabilitytheory1971}%
  \BibitemOpen
  \bibfield  {author} {\bibinfo {author} {\bibfnamefont {W.}~\bibnamefont
  {Feller}},\ }\href@noop {} {\emph {\bibinfo {title} {An Introduction to
  Probability Theory and Its Applications}}}\ (\bibinfo  {publisher}
  {{Wiley}},\ \bibinfo {year} {1971})\BibitemShut {NoStop}%
\bibitem [{\citenamefont {Lekner}(2016)}]{LeknerTightfocusinglight2016}%
  \BibitemOpen
  \bibfield  {author} {\bibinfo {author} {\bibfnamefont {J.}~\bibnamefont
  {Lekner}},\ }\bibfield  {title} {\enquote {\bibinfo {title} {Tight focusing
  of light beams: A set of exact solutions},}\ }\href {\doibase
  10.1098/rspa.2016.0538} {\bibfield  {journal} {\bibinfo  {journal}
  {Proceedings of the Royal Society A: Mathematical, Physical and Engineering
  Science}\ }\textbf {\bibinfo {volume} {472}},\ \bibinfo {pages} {20160538}
  (\bibinfo {year} {2016})}\BibitemShut {NoStop}%
\bibitem [{\citenamefont {Andrejić}\ and\ \citenamefont
  {Lekner}(2017)}]{Andrejic2017}%
  \BibitemOpen
  \bibfield  {author} {\bibinfo {author} {\bibfnamefont {P.}~\bibnamefont
  {Andrejić}}\ and\ \bibinfo {author} {\bibfnamefont {J.}~\bibnamefont
  {Lekner}},\ }\bibfield  {title} {\enquote {\bibinfo {title} {Topology of
  phase and polarisation singularities in focal regions},}\ }\href {\doibase
  10.1088/2040-8986/aa895d} {\bibfield  {journal} {\bibinfo  {journal} {Journal
  of Optics}\ }\textbf {\bibinfo {volume} {19}},\ \bibinfo {pages} {105609}
  (\bibinfo {year} {2017})}\BibitemShut {NoStop}%
\bibitem [{\citenamefont {Lekner}\ and\ \citenamefont
  {Andrejić}(2018)}]{Lekner2018}%
  \BibitemOpen
  \bibfield  {author} {\bibinfo {author} {\bibfnamefont {J.}~\bibnamefont
  {Lekner}}\ and\ \bibinfo {author} {\bibfnamefont {P.}~\bibnamefont
  {Andrejić}},\ }\bibfield  {title} {\enquote {\bibinfo {title} {Nonexistence
  of exact solutions agreeing with the gaussian beam on the beam axis or in the
  focal plane},}\ }\href {\doibase 10.1016/j.optcom.2017.08.071} {\bibfield
  {journal} {\bibinfo  {journal} {Optics Communications}\ }\textbf {\bibinfo
  {volume} {407}},\ \bibinfo {pages} {22--26} (\bibinfo {year}
  {2018})}\BibitemShut {NoStop}%
\bibitem [{\citenamefont {Watson}(1922)}]{WatsonTreatise1922}%
  \BibitemOpen
  \bibfield  {author} {\bibinfo {author} {\bibfnamefont {G.~N.}\ \bibnamefont
  {Watson}},\ }\href@noop {} {\emph {\bibinfo {title} {A Treatise on the Theory
  of Bessel Functions (Cambridge Mathematical Library)}}}\ (\bibinfo
  {publisher} {Cambridge University Press},\ \bibinfo {year}
  {1922})\BibitemShut {NoStop}%
\bibitem [{\citenamefont {Dorn}, \citenamefont {Quabis},\ and\ \citenamefont
  {Leuchs}(2003)}]{DornSharperFocusRadially2003}%
  \BibitemOpen
  \bibfield  {author} {\bibinfo {author} {\bibfnamefont {R.}~\bibnamefont
  {Dorn}}, \bibinfo {author} {\bibfnamefont {S.}~\bibnamefont {Quabis}}, \ and\
  \bibinfo {author} {\bibfnamefont {G.}~\bibnamefont {Leuchs}},\ }\bibfield
  {title} {\enquote {\bibinfo {title} {Sharper {{Focus}} for a {{Radially
  Polarized Light Beam}}},}\ }\href {\doibase 10.1103/PhysRevLett.91.233901}
  {\bibfield  {journal} {\bibinfo  {journal} {Phys. Rev. Lett.}\ }\textbf
  {\bibinfo {volume} {91}},\ \bibinfo {pages} {233901} (\bibinfo {year}
  {2003})}\BibitemShut {NoStop}%
\bibitem [{\citenamefont {Quabis}\ \emph {et~al.}(2000)\citenamefont {Quabis},
  \citenamefont {Dorn}, \citenamefont {Eberler}, \citenamefont {Gl{\"o}ckl},\
  and\ \citenamefont {Leuchs}}]{QuabisFocusinglighttighter2000a}%
  \BibitemOpen
  \bibfield  {author} {\bibinfo {author} {\bibfnamefont {S.}~\bibnamefont
  {Quabis}}, \bibinfo {author} {\bibfnamefont {R.}~\bibnamefont {Dorn}},
  \bibinfo {author} {\bibfnamefont {M.}~\bibnamefont {Eberler}}, \bibinfo
  {author} {\bibfnamefont {O.}~\bibnamefont {Gl{\"o}ckl}}, \ and\ \bibinfo
  {author} {\bibfnamefont {G.}~\bibnamefont {Leuchs}},\ }\bibfield  {title}
  {\enquote {\bibinfo {title} {Focusing light to a tighter spot},}\ }\href
  {\doibase 10.1016/S0030-4018(99)00729-4} {\bibfield  {journal} {\bibinfo
  {journal} {Optics Communications}\ }\textbf {\bibinfo {volume} {179}},\
  \bibinfo {pages} {1--7} (\bibinfo {year} {2000})}\BibitemShut {NoStop}%
\bibitem [{\citenamefont {Quabis}\ \emph {et~al.}(2001)\citenamefont {Quabis},
  \citenamefont {Dorn}, \citenamefont {Eberler}, \citenamefont {Gl{\"o}ckl},\
  and\ \citenamefont {Leuchs}}]{Quabisfocuslighttheoretical2001}%
  \BibitemOpen
  \bibfield  {author} {\bibinfo {author} {\bibfnamefont {S.}~\bibnamefont
  {Quabis}}, \bibinfo {author} {\bibfnamefont {R.}~\bibnamefont {Dorn}},
  \bibinfo {author} {\bibfnamefont {M.}~\bibnamefont {Eberler}}, \bibinfo
  {author} {\bibfnamefont {O.}~\bibnamefont {Gl{\"o}ckl}}, \ and\ \bibinfo
  {author} {\bibfnamefont {G.}~\bibnamefont {Leuchs}},\ }\bibfield  {title}
  {\enquote {\bibinfo {title} {The focus of light \textendash{} theoretical
  calculation and experimental tomographic reconstruction},}\ }\href {\doibase
  10.1007/s003400000451} {\bibfield  {journal} {\bibinfo  {journal} {Applied
  Physics B}\ }\textbf {\bibinfo {volume} {72}},\ \bibinfo {pages} {109--113}
  (\bibinfo {year} {2001})}\BibitemShut {NoStop}%
\bibitem [{\citenamefont {Kozawa}\ and\ \citenamefont
  {Sato}(2007)}]{KozawaSharperfocalspot2007}%
  \BibitemOpen
  \bibfield  {author} {\bibinfo {author} {\bibfnamefont {Y.}~\bibnamefont
  {Kozawa}}\ and\ \bibinfo {author} {\bibfnamefont {S.}~\bibnamefont {Sato}},\
  }\bibfield  {title} {\enquote {\bibinfo {title} {Sharper focal spot formed by
  higher-order radially polarized laser beams},}\ }\href {\doibase
  10.1364/JOSAA.24.001793} {\bibfield  {journal} {\bibinfo  {journal} {J. Opt.
  Soc. Am. A, JOSAA}\ }\textbf {\bibinfo {volume} {24}},\ \bibinfo {pages}
  {1793--1798} (\bibinfo {year} {2007})}\BibitemShut {NoStop}%
\bibitem [{\citenamefont {Arkhipov}\ and\ \citenamefont
  {Schulten}(2009)}]{ArkhipovLimitsreductioneffective2009}%
  \BibitemOpen
  \bibfield  {author} {\bibinfo {author} {\bibfnamefont {A.}~\bibnamefont
  {Arkhipov}}\ and\ \bibinfo {author} {\bibfnamefont {K.}~\bibnamefont
  {Schulten}},\ }\bibfield  {title} {\enquote {\bibinfo {title} {Limits for
  reduction of effective focal volume in multiple-beam light microscopy},}\
  }\href {\doibase 10.1364/OE.17.002861} {\bibfield  {journal} {\bibinfo
  {journal} {Opt. Express, OE}\ }\textbf {\bibinfo {volume} {17}},\ \bibinfo
  {pages} {2861--2870} (\bibinfo {year} {2009})}\BibitemShut {NoStop}%
\bibitem [{\citenamefont {Olver}\ and\ \citenamefont {{National Institute of
  Standards and Technology (U.S.)}}(2010)}]{OlverNISThandbookmathematical2010}%
  \BibitemOpen
  \bibinfo {editor} {\bibfnamefont {F.~W.~J.}\ \bibnamefont {Olver}}\ and\
  \bibinfo {editor} {\bibnamefont {{National Institute of Standards and
  Technology (U.S.)}}},\ eds.,\ \href@noop {} {\emph {\bibinfo {title}
  {{{NIST}} Handbook of Mathematical Functions}}}\ (\bibinfo  {publisher}
  {{Cambridge University Press : NIST}},\ \bibinfo {address} {Cambridge ; New
  York},\ \bibinfo {year} {2010})\BibitemShut {NoStop}%
\bibitem [{\citenamefont {Jeffrey}\ and\ \citenamefont
  {Zwillinger}(2007)}]{JeffreyTableIntegralsSeries2007}%
  \BibitemOpen
  \bibfield  {author} {\bibinfo {author} {\bibfnamefont {A.}~\bibnamefont
  {Jeffrey}}\ and\ \bibinfo {author} {\bibfnamefont {D.}~\bibnamefont
  {Zwillinger}},\ }\href@noop {} {\emph {\bibinfo {title} {Table of
  {{Integrals}}, {{Series}}, and {{Products}}}}},\ \bibinfo {edition} {7th}\
  ed.\ (\bibinfo  {publisher} {{Elsevier}},\ \bibinfo {year}
  {2007})\BibitemShut {NoStop}%
\end{thebibliography}%
\newpage
\onecolumngrid{}

\appendix
\section{Identities}\label{app:identities}
The integral of a beam profile over a plane $z=z_0$ can be expressed as the integral over its weight functions in Bessel beam representation. We can write the beam wavefunction $\psi$ as
\begin{equation}
  \psi(\rho,\phi,z_0)=\frac{A_0}{k^2}\sum_{m\in\mathbb{Z}} \ee^{\ii (m \phi+q(\kappa)z_0)}\int_0^k f_m(\kappa/k)J_m(\kappa\rho)\kappa\du{\kappa}, \quad q(\kappa)=\sqrt{k^2-\kappa^2}.
\end{equation}
The integral over the plane of the modulus squared is given by
\begin{equation}
\begin{aligned}
  \int_{z=z_0} \modu[2]{\psi}\du[2]{r} =
 &\frac{\ma}{k^4}\int_0^{2\pi}\int_0^\infty 
  \sum_{m,m'\in\mathbb{Z}} \ee^{\ii \left((m-m')\phi+(q(\kappa)-q(\kappa'))z_0\right)}
  \\
  \times & \iint_0^k f_m(\kappa/k){f_{m'}(\kappa'/k)}^* J_m(\kappa\rho)J_{m'}(\kappa'\rho)\kappa\du{\kappa}\kappa'\du{\kappa'}\rho\du{\rho}\du{\phi}.
\end{aligned}
\end{equation}
The relevant orthogonality relations are
\begin{equation}
  \begin{gathered}
    \int_0^\infty J_m(\kappa\rho)J_m(\kappa'\rho)\rho\du{\rho}=\frac{\delta(\kappa-\kappa')}{\kappa-\kappa'},
    \\
    \int_0^{2\pi}\ee^{\ii(m-m')\phi}\du{\phi}=2\pi\delta_{m,m'}.
  \end{gathered}
\end{equation}
Applying the two relations gives the simple expression
\begin{equation}\label{eq:int-all-space}
    \int_{z=z_0} \modu[2]{\psi}\du[2]{r} =\frac{2\pi \modu[2]{A_0}}{k^4} \sum_{m\in\mathbb{Z}}\int_0^k \modu[2]{f_m(\kappa/k)}\kappa\du{\kappa}.
\end{equation}
Note that this is independent of $z_0$, which is only true if the wave has all forward propagating plane wave components.
Evaluating the wave-function at the origin has $J_m(0)=0$ unless $m=0$, in which case we have $J_0(0)=1$. Therefore the value at the origin is simply the integral over the zeroth weight function,
\begin{equation}\label{eq:int-origin}
  \psi_0=\psi\big|_{\rho=0}=\frac{A_0}{k^2}\int_0^k f_0(\kappa/k)\kappa\du{\kappa}.
\end{equation}

\section{Radial integral of plane waves}\label{app:radial-integral}
We can write a plane wave of arbitrary direction $\ee^{\ii \vv{k}\cdot\vv{r}}$ in cylindrical spatial coordinates as
\begin{equation}
	\ee^{\ii \vv{k}\cdot\vv{r}}=\exp\left(\ii \left[kz\cos\alpha+k\rho \sin\alpha \cos(\phi-\beta)\right]\right).
\end{equation}
The generating function for Bessels is (NIST Handbook~\cite{OlverNISThandbookmathematical2010})
\begin{equation}
	\exp\left(\frac{1}{2}\zeta\left[\tau-\tau^{-1}\right]\right)=\sum_{m=-\infty}^{\infty}\tau^m J_m(\zeta).
\end{equation}
We can set $\zeta=k\rho\sin\alpha$, $\tau=i\ee^{\ii m(\phi-\beta)}$ to get
\begin{equation}
	\ee^{\ii k\rho\sin\cos(\phi-\beta)}=\sum_{m=-\infty}^{\infty}\ii^m \ee^{\ii m(\phi-\beta)}J_m(k\rho\sin\alpha).
\end{equation}
The radial integral of $\ee^{\ii (\vv{k}-\vv{k}')\cdot\vv{r}}$ gives
\begin{equation}
\begin{gathered}
	\int_{0}^{\infty}\rho\du{\rho}\int_{0}^{2\pi}\du{\phi}\ee^{\ii (\vv{k}-\vv{k}')\cdot\vv{r}}
	\\
	=\ee^{ikz(\cos\alpha-\cos\alpha')}\int_{0}^{\infty}\rho\du{\rho}\int_{0}^{2\pi}\sum_{m,m'=-\infty}^{\infty}
	\ii^{m-m'}\ee^{\ii(m-m')\phi}\ee^{\ii (m'\beta'-m\beta)}J_m(k\rho\sin\alpha)J_m(k\rho\sin\alpha')
	\\
	=2\pi\sum_{m=-\infty}^{\infty}\ee^{\ii m(\beta-\beta')}\frac{\delta(k\sin\alpha-k\sin\alpha')}{k\sin\alpha}
	\\
	=
	\frac{4\pi^2\delta(\beta-\beta')\delta(\alpha-\alpha')}{k^2\sin\alpha\cos\alpha}
\end{gathered}
\end{equation}
Here we have used the fact that~\cite[eq. 1.17.21]{OlverNISThandbookmathematical2010}
\begin{equation}
	\sum_{m=-\infty}^{\infty}\ee^{\ii m(\beta-\beta')}=2\pi\delta(\beta-\beta')
\end{equation}
is a representation of the Dirac delta function on the interval $[0,2\pi]$, and we have used the functional composition property of the Dirac delta to rewrite
\begin{equation}
	\delta(k\sin\alpha-k\sin\alpha')=\frac{\delta(\alpha-\alpha')}{k\cos\alpha}.
\end{equation}
Here, I am assuming that the range of $\alpha,\alpha'$ is $[0,\pi/2]$, with $\sin\alpha$ has only one zero on this interval. In Section~\ref{sec:electromagnetic} I consider forward propagating waves only, so this assumption holds.

\section{Focal concentration for Laguerre Gauss beams}\label{app:focal-concentration}
All factors independent of $\rho$ in the definition of the Laguerre-Gauss beam in~\eqref{eq:laguerre-gauss} can be neglected for this calculation, as they will cancel when taking the ratio of integrals. The quantity of interest is the intensity per unit area, which has $\rho$ dependence proportional to
\begin{equation}
  I_{p,m}(\rho,z)=\modu[2]{A_0}{\left(\frac{\rho^2}{{W_G}^2}\right)}^{|m|}{\left(L_p^{|m|}\left(\frac{\rho^2}{{W_G}^2}\right)\right)}^2\exp\left(-\frac{\rho^2}{{W_G}^2}\right).
\end{equation}
This can be simplified by making the substitution $\xi =\rho^2/{W_G}^2$. We also have $\zeta=s^2/{W_G}^2$ for our regulator length, and $\rho\du{\rho}={W_G}^2\du{\xi}/2$. The focal concentration is therefore given by
\begin{equation}
  \alpha_{p,m}(s)=\frac{\pi s^2 \int I_{p,m}\rho\du{\rho}}{\int I_{p,m}\ee^{-\rho^2/s^2}\rho\du{\rho}}=\pi {W_G}^2\frac{\int I_{p,m}\du{\xi}}{\int I_{p,m}\ee^{-\xi/\zeta}\du{\xi}}.
\end{equation}
The numerator is proportional to~\cite[p.~809, eq.~7.414.3]{JeffreyTableIntegralsSeries2007}
\begin{equation}\label{eq:num}
  \int_0^\infty \xi^{|m|}\left(L_p^{|m|}(\xi)\right)^2\ee^{-\xi}\du{\xi}=\frac{(p+|m|)!}{p!}
\end{equation}
The denominator is more complex, and is proportional to~\cite[p.~809, eq.~7.414.4]{JeffreyTableIntegralsSeries2007}
\begin{multline}\label{eq:den}
  \int_0^\infty \xi^{|m|}
  \left(L_p^{|m|}(\xi)\right)^2
  \ee^{-\beta\xi}\du{\xi}
  =\frac{(2p+|m|)!}{{(p!)}^2}\frac{{(\beta-1)}^{2p}}{\beta^{2p+|m|+1}}\\
  \times {}_2F_1\left(-p,-p;-2p-|m|,\frac{\beta(\beta-2)}{{(\beta-1)}^2}\right), \quad \beta\equiv 1+\frac{1}{\zeta}.
\end{multline}
Upon substituting $\beta=1-\frac{1}{\zeta}$, and $\zeta=s^2/W_G^2$, and taking the ratio of~\eqref{eq:num} over~\eqref{eq:den}, one arrives at the expression
\begin{equation}
  \alpha_{p,m}(s)=\alpha_G(s)\frac{\left(\frac{s}{W_G}\right)^{4|m|}p!  (p+|m|)! \left(\frac{{W_G}^2}{s^2}+1\right)^{2p+|m|}}{(2p+|m|)! \, _2F_1\left(-p,-p;-2p-|m|;1-\frac{s^4}{{W_G}^4}\right)},\quad\alpha_G(s)=\pi(s^2+{W_G}^2).
\end{equation}

\end{document}